\documentclass[aps,prl,twocolumn,showpacs,superscriptaddress,groupedaddress]{revtex4}  
\usepackage{graphicx}  
\usepackage{color}
\usepackage{bm}        
\usepackage{amssymb}   
\begin{document}
\title{Path-dependent Entropy Production}
\author{Chulan Kwon}
\affiliation{Department of Physics, Myongji University, Yongin, Gyeonggi-Do, 17058, Korea}
\email{ckwon@mju.ac.kr}

\date{\today}

\begin{abstract}
A rigorous derivation of nonequilibrium entropy production  via the path-integral formalism is presented. Entropy production is defined as the entropy change piled in a heat reservoir as a result of a nonequilibrium thermodynamic process. It is a central quantity by which various forms of the fluctuation theorem are obtained.  The two kinds of the stochastic dynamics are investigated: the Langevin dynamics for an even-parity state and the Brownian motion of a single particle.  Mathematical ambiguities in deriving the functional form of the entropy production, which depends on path in state space, are clarified by using a rigorous quantum mechanical approach.
\end{abstract}



\maketitle
\section{Introduction}
The fluctuation theorem (FT) is an important principle in nonequilibrium statistical mechanics. It was first discovered two decades ago for a deterministic molecular dynamics~\cite{evans}, but was later proven to hold for stochastic dynamics~\cite{jarzynski,crooks,kurchan,lebowitz}. It deals with thermal fluctuations of time-accumulated quantities such as work, heat, and entropy production (EP). In the usual statistical-mechanics approach, only average quantities are of interest because fluctuations are negligible in the thermodynamic limit. However, owing to modern technologies, one can investigate the thermal motion of a small system on a nano-scale that shows large fluctuations. For example, one can observe that the thermodynamic law is violated in rare events occurring due to extreme fluctuations. The FT provides a rigorous relation for the probability distribution function of such a fluctuating quantity and has been proven for various small-system experiments~\cite{wang,trepagnier,hummer,garnier,douarche,joubaud,liphardt,collin,hayashi,pak}.

The path-integral theory for the thermal motion was developed by Onsager and Machlup in 1953~\cite{onsager} and has been used extensively to study the fluctuation theorem and related issues. Finding the EP from the path probability is essential in order to prove the FT.
Consider a particle evolving along a path $\mathbf{q}(t)$ in state space under a time-dependent protocol $\lambda(t)$ for time period $0<t<\tau$, starting from $\mathbf{q}_0$ and going to $\mathbf{q}_{\tau}$. The state vector $\mathbf{q}$ may be the position $\mathbf{x}$ in an overdamped case or the position-momentum pair $(\mathbf{x},\mathbf{p})$ in an underdamped case. Infinitely many paths connect the initial and the final states,  among which a single path is determined stochastically from the given dynamics. 
Given a path $\mathbf{q}(t)$, the time-reverse path is given by
\begin{equation}
\bar{\mathbf{q}}(t)=\left\{
\begin{array}{ll}
\mathbf{x}(\tau-t)&\textrm{(even parity)} \\
(\mathbf{x}(\tau-t),-\mathbf{p}(\tau-t)) & \textrm{(mixture of even and odd parity)}
\end{array}~,
\right.
\end{equation}
where $\mathbf{x}$ has even parity and 
the momentum $\mathbf{p}$ has odd parity in time reversal $t\to\tau-t$.
Let $\Pi[\mathbf{q}(t);\lambda(t)]$ be the conditional probability for the system to evolve along the forward (given) path under the protocol and 
$\Pi[\bar{\mathbf{q}}(t);\bar{\lambda}(t)]$ be that for the time-reverse path under the time-reverse protocol $\bar{\lambda}(t)=\lambda(\tau-t)$. Then, the EP is given by
\begin{equation}
\Delta S_{\textrm{env}}=k_B\ln \frac{\Pi[\mathbf{q}(t);\lambda(t)]}{ \Pi[\bar{\mathbf{q}}(t);\bar{\lambda}(t)]}~,
\label{detailed_FR}
\end{equation}
where $\Delta S_{\textrm{env}}$ is another name for the EP denoting the entropy change in the environment (heat reservoir), and $k_B$ is the Boltzmann constant.
This relation is called {\em the detailed fluctuation relation}~\cite{crooks} in the literature, which was first derived by Schnackenberg for the master equation for discrete states~\cite{schnakenberg}.

Then, the change in the total entropy of the system and the heat reservoir is given by
\begin{equation}
\Delta S_{\textrm{tot}}=k_B\ln \frac{\rho(\mathbf{q}_0)\Pi[\mathbf{q}(t);\lambda(t)]}{ \rho(\mathbf{q}_{\tau})\Pi[\bar{\mathbf{q}}(t);\bar{\lambda}(t)]}~,
\end{equation}
where $\rho(\mathbf{q}_0)$ are $\rho(\mathbf{q}_{\tau})$ are the initial and the final probability distribution functions (PDF), respectively. Recognizing the system entropy as the Shannon entropy, $-\ln\rho(\mathbf{x})$~\cite{shannon}, the change in the system entropy is given by $\Delta S_{\textrm{sys}}=k_B[\ln\rho(\mathbf{q}_0)-\rho(\mathbf{q}_{\tau})]$, so $\Delta S_{ \textrm{tot}}=\Delta S_{\textrm{sys}}+\Delta S_{ \textrm{env}}~.$ Then, the FT for the total entropy change can be easily proven to be
\begin{eqnarray}
\langle e^{-k_B^{-1}\Delta S_{\textrm{tot}}}\rangle &=& \int d\mathbf{q}_0\rho(\mathbf{q}_0)
\int D[\mathbf{q}(t)]\Pi[\mathbf{q}(t);\lambda(t)]\nonumber\\
&&\times \frac{ \rho(\mathbf{q}_{\tau})\Pi[\bar{\mathbf{q}}(t);\bar{\lambda}(t)]}{\rho(\mathbf{q}_0)\Pi[\mathbf{q}(t);\lambda(t)]}\\
\label{FT_EP}
&=&\int d\bar{\mathbf{q}}_0\rho(\mathbf{q}_{\tau})
\int D[\bar{\mathbf{q}}(t)]\Pi[\bar{\mathbf{q}}(t);\bar{\lambda}(t)]=1\nonumber~,
\end{eqnarray}
where the Jacobian for the change of variable $\mathbf{q}\to\bar{\mathbf{q}}$ is equal to 1 and $\rho(\mathbf{q}_{\tau})$ is the initial PDF for the reverse path.
Using Jensen' s inequality, one can get $\langle\Delta S_{\textrm{tot}}\rangle\ge 0$, which confirms the second law of thermodynamics. Other types of FT's can also be proven in similar ways where the use of the detailed fluctuation relation in Eq.~(\ref{detailed_FR}) is crucial~\cite{seifert}.
 
In this paper, the detailed fluctuation relations for two kinds of the stochastic dynamics are derived: (i) the Langevin dynamics for an even-parity state and (ii) the Brownian motion of a single particle.  The dynamics (i) is described by
\begin{equation}
\dot{\mathbf{x}}=\mathbf{g}(\mathbf{x},\lambda(t),t)+\boldsymbol{\xi}(t)~,\label{overdamped}
\end{equation}
where $\boldsymbol{\xi}$ is the noise vector  with zero mean and with the variance  given by $\langle\boldsymbol{\xi}(t)\boldsymbol{\xi}(t')\rangle=2\mathsf{D}\delta(t-t')$. The outer product $\mathbf{c}\mathbf{d}$ for vectors $\mathbf{c}$ and $\mathbf{d}$ denotes a matrix with element $c_md_n$ for the $m^{\textrm{th}}$ row and the $n^{\textrm{th}}$ column. $\mathsf{D}$ is called the diffusion matrix, and is symmetric and positive-definite. $\mathbf{g}$ is the drift term, is proportional to the applied force in the overdamped limit, and depends on the protocol as well as the state of the system. 
The dynamics (ii) is described by 
\begin{equation}
\left\{
\begin{array}{l}
\dot{\mathbf{x}}=\mathbf{p}/m \\
\dot{\mathbf{p}}=-\mathsf{G}\cdot\mathbf{p}/m+\mathbf{f}(\mathbf{x},
\mathbf{p},\lambda(t),t)+\boldsymbol{\zeta}(t)
\end{array}\right.~,\label{underdamped}
\end{equation}
where $\mathbf{x}$ and $\mathbf{p}$ are the position and momentum of the particle, $\mathsf{G}$ the friction matrix, and $m$ the mass of the particle. $\mathbf{f}$ is the applied force on the system, which generally depends on the position and the momentum as well as the protocol. $\mathsf{S}$ is the diffusion matrix for the noise vector $\boldsymbol{\zeta}$,  given from $\langle\boldsymbol{\zeta}(t)\boldsymbol{\zeta}(t')\rangle=2\mathsf{S}\delta(t-t')$. In the overdamped limit with large $\mathsf{G}$ or equivalently in the zero-mass limit, (ii) is found to reduce to (i) if $\mathbf{f}$ is independent of $\mathbf{p}$. In reality, the momentum-dependence of $\mathbf{f}$ can be found in active matters~\cite{active,granular,active_Brown} and the feedback control~\cite{cold_damping,khkim,Chulan_information}. Many unprecedented features have recently been found for such an odd-parity force~\cite{spinney,hklee,kwon-yeo-lee-park}, but not for any even-parity force. Only non-multiplicative noise where the diffusion matrix is independent of $\mathbf{x}$ or $\mathbf{p}$ is considered, and a study of the multiplicative case will be presented in a separate place. 

\section{Langevin dynamics for an even-parity state}
The differential equation for the PDF associated with the Langevin equation, Eq.~(\ref{overdamped}), is given by the Fokker-Planck equation~\cite{risken}:
\begin{equation}
\partial_t \rho(\mathbf{x}, t)=-H_{\textrm{FP}}\rho(\mathbf{x}, t)~,
\label{FP_equation}
\end{equation}
where $\partial_t=\partial/\partial t$. $H_{\textrm{FP}}$ is a non-Hermitian operator given by
\begin{equation}
H_{\textrm{FP}}=\partial_{\mathbf{x}}\cdot\left[\mathbf{g}-\mathsf{D}\cdot
\partial_{\mathbf{x}}\right]~,
\label{FP_operator}
\end{equation}
where $\partial_{\mathbf{x}}=\boldsymbol{\nabla}_{\mathbf{x}}$.
Introducing an operator $\hat{\mathbf{a}}=i^{-1}\partial_{\mathbf{x}}$ for $i=\sqrt{-1}$ conjugate to $\mathbf{x}$, one can write
\begin{equation}
H_{\textrm{FP}} =i\hat{\mathbf{a}}\cdot\mathbf{g}+\hat{\mathbf{a}}\cdot \mathsf{D}\cdot\hat{\mathbf{a}}~.
\end{equation}
Using the quantum-mechanics formalism, $\rho(\mathbf{x}, t)=\langle \mathbf{x}|\rho(t)\rangle=
\langle\mathbf{x}|e^{-t H_{\textrm{FP}}}|\mathbf{x}_{0}\rangle\langle\mathbf{x}_0|\rho(0)\rangle$, where the integration over $\mathbf{x}_0$ is implicit following the Einstein convention. The propagator $\langle\mathbf{x}|e^{-t H_{\textrm{FP}}}|\mathbf{x}_{0}\rangle$ in the quantum-mechanical sense turns out be the conditional probability $P(\mathbf{x},t|\mathbf{x}_0,0)$ where stochasticity (normalizability) is guaranteed by $\int d\mathbf{x} \langle\mathbf{x}|H_{\textrm{FP}}|\mathbf{x}_{0}\rangle=0$, which can be seen easily from the total derivative in Eq.~(\ref{FP_operator}). 

For $0<t<\tau$, let us discretize time by by using the interval $\Delta t=\tau/N$ as $t_j=j\Delta t$ for integer $j$. For $0\le j\le N$, $t_0=0$, $t_N=\tau$. In the $N\to\infty$ ($\Delta t\to 0$) limit, one can write  
\begin{equation}
P(\mathbf{x}_{\tau},\tau|\mathbf{x}_0,0)=\int \left[\prod_{j=1}^{N-1} d\mathbf{x}_j\right]\prod_{j=1}^{N}\langle\mathbf{x}_j|e^{-\Delta t H_{\textrm{FP}}}|\mathbf{x}_{j-1}\rangle~,
\end{equation}
where $\mathbf{x}_j=\mathbf{x}(t_j)$. A set $\{\mathbf{x}_0,\mathbf{x}_1,\ldots,\mathbf{x}_j,\ldots,\mathbf{x}_N\}$ is a discrete representation of a path $\mathbf{x}(t)$. The integration over all intermediate states $\mathbf{x}_j$ for $1\le j\le N-1$ corresponds to the integration over all paths connecting $\mathbf{x}_0$ and $\mathbf{x}_{\tau}$. Then, one can identify the path probability, $\Pi[\mathbf{x}(t);\lambda(t)]$, as $\prod_{j=1}^{N}\langle\mathbf{x}_j|e^{-\Delta t H_{\textrm{FP}}}|\mathbf{x}_{j-1}\rangle$, where $\lambda(t)$ takes the value $\lambda_{j-1}$ at time $t_{j-1}$ for the interval between $t_{j-1}$ and $t_j$.

One can change the order of $\mathbf{a}$ in $H_{\textrm{FP}}$ as  $i\hat{\mathbf{a}}\cdot\mathbf{g}=i\mathbf{g}\cdot \hat{\mathbf{a}}+\partial_{\mathbf{x}}\cdot \mathbf{g}$, by using the commutator $[\hat{\mathbf{a}}, \mathbf{g}]=i^{-1}\partial_{\mathbf{x}} \mathbf{g}$. Changing the order partially, one can write
\begin{equation}
H_{\textrm{FP}}= (1-\alpha)H_{\textrm{FP}}+\alpha \tilde{H}_{\textrm{FP}}~,
\end{equation}
where $0\le \alpha\le 1$ and $\tilde{H}_{\textrm{FP}}=i\mathbf{g}\cdot\hat{\mathbf{a}}+\partial_{\mathbf{x}}\cdot\mathbf{g}+\hat{\mathbf{a}}
\cdot\mathsf{D}\cdot\hat{\mathbf{a}}$. For eigenstate $|\mathbf{a}\rangle$ of $\hat{\mathbf{a}}$, $1=|\mathbf{a}\rangle\langle\mathbf{a}|$ and $\langle\mathbf{x}|\mathbf{a}\rangle=\langle\mathbf{a}|\mathbf{x}\rangle^*=e^{i\mathbf{a}\cdot\mathbf{x}}/(2\pi)^{d/2}$ in $d$-dimensional space. Then, one can find the expression in detail for the propagator between $t_{j-1}$ and $t_j$ as 
\begin{eqnarray}
\lefteqn{\langle\mathbf{x}_j|e^{-\Delta t H_{\textrm{FP}}}|\mathbf{x}_{j-1}\rangle}\nonumber\\
&\simeq&\langle\mathbf{x}_j| \mathbf{a}\rangle\langle \mathbf{a}|\left[1-\Delta t(1-\alpha)H_{\textrm{FP}} \right]|\mathbf{x}_{j-1}\rangle\nonumber\\
&&-\langle\mathbf{x}_j|\alpha\Delta t \tilde{H}_{\textrm{FP}}| \mathbf{a}\rangle\langle \mathbf{a}|\mathbf{x}_{j-1}\rangle\nonumber\\
&\simeq&\int \frac{d\mathbf{a}}{(2\pi)^d}
e^{-\Delta t\mathbf{a}\cdot\mathsf{D}\cdot\mathbf{a}
+i\mathbf{a}\cdot(\mathbf{x}_j-\mathbf{x}_{j-1})-i\Delta t\mathbf{a}\cdot \mathbf{g}_j^{(\alpha)}-\Delta t\alpha(\partial_{\mathbf{x}}\cdot\mathbf{g})_j}\nonumber\\
&=&\frac{1}{[\textrm{det}(4\pi\Delta t\mathsf{D})]^{1/2}}
e^{-\frac{\Delta t}{4}\mathbf{h}_j^{(\alpha)}\cdot\mathsf{D}^{-1}\cdot\mathbf{h}_j^{(\alpha)}-\Delta t\alpha (\partial_{\mathbf{x}}\cdot\mathbf{g})_j}~,
\label{forward_segment}
\end{eqnarray}
where $\mathbf{h}_j^{(\alpha)}=
(\mathbf{x}_j-\mathbf{x}_{j-1})/\Delta t-\mathbf{g}_{j}^{(\alpha)}$. $\mathbf{g}_{j}^{(\alpha)}$ is an intermediate value between $\mathbf{g}$ at $\mathbf{x}_{j-1}$ and $\mathbf{x}_{j}$ and is defined by $(1-\alpha)\mathbf{g}_{j-1}+\alpha\mathbf{g}_{j}$ for $\mathbf{g}_k=\mathbf{g}(\mathbf{x}_k,\lambda_{j-1})$ for both $k=j-1$ and $j$. Special cases are $\alpha=0$ (prepoint), $\alpha=1/2$ (midpoint), and $\alpha=1$ (postpoint). The meanings of the terms in the brackets are clear. Note that an arbitrary use of $\alpha$ for the path integral leads to the same result up to higher-order correction $\mathcal{O}(\Delta t)$. 

Then, the path probability is found as
\begin{eqnarray}
\lefteqn{\Pi[\mathbf{x}(t);\lambda(t)]}\nonumber\\
&=&\lim_{N\to\infty}\mathcal{N}
e^{-\sum_{j=1}^{N}\Delta t\left[\frac{1}{4}\mathbf{h}_j^{(\alpha)}\cdot\mathsf{D}^{-1}\cdot\mathbf{h}_j^{(\alpha)}+\alpha (\partial_{\mathbf{x}}\cdot\mathbf{g})_j\right]}\nonumber\\
&=&\mathcal{N}e^{-\int_0^{\tau} dt\left[\frac{1}{4}\mathbf{h}(\mathbf{x},\dot{\mathbf{x}},t)\cdot\mathsf{D}^{-1}\cdot\mathbf{h}(\mathbf{x},\dot{\mathbf{x}},t)+\alpha \partial_{\mathbf{x}}\cdot\mathbf{g}\right]}~,
\label{forward}
\end{eqnarray}
where $\mathcal{N}=[\textrm{det}(4\pi\Delta t\mathsf{D})]^{-N/2}$and $\mathbf{h}(\mathbf{x},\dot{\mathbf{x}},t)=\dot{\mathbf{x}}-\mathbf{g}$. The last line is written in the continuous-time limit  while it is only well-defined in the discrete-time representation. The integration of Eq.~(\ref{forward}) over all paths and over the initial state, given the initial PDF, goes to 1, which can be written as $1=\int d\mathbf{x}_0\rho(\mathbf{x}_0)\int D[\mathbf{x}(t)] \Pi[\mathbf{x}(t);\lambda(t)]$. This normalization property is used in Eq.~(\ref{FT_EP}), where $D[\mathbf{x}(t)]=\lim_{N\to\infty}\mathcal{N}\prod_{j=1}^{N}d\mathbf{x}_j$.
For the prepoint representation, the continuum limit can be seen to be the same as the noise distribution by noting $\dot{x}-\mathbf{g}=\boldsymbol{\xi}(t)$, agreeing with common intuition. The seeming ambiguity of the path integral, particularly in treating the divergence term $\partial_{\mathbf{x}}\cdot\mathbf{g}$, can be well resolved by a correct use of the $\alpha$-representation in Eq.~(\ref{forward}), the choice of which is a matter of taste. The path integral can be exactly carried out when $\mathbf{g}$ is linear in $\mathbf{x}$, which was done for the linear diffusion problem using the prepoint representation~\cite{kwon_2011}. It was also done using the midpoint representation for a similar linear problem of the injected and the dissipated powers~\cite{jslee}. 

The time-reverse path is given by $\{0\le j\le N|\bar{\mathbf{x}}_j=\mathbf{x}_{N-j}\}$. For  the time interval between $t_{j-1}$ and $t_{j}$, the drift term is denoted as $\bar{\mathbf{g}}_k=\mathbf{g}(\bar{\mathbf{x}}_k,\bar{\lambda}_{j})$ for both $k=j-1$ and $j$, where the protocol has a fixed value at the later time, opposite to the forward path. This is the strategy for the dynamics to have a common protocol for an overlapping segment of path. Replacing $j$ with $N-j+1$, the transition occurs reversely from $\mathbf{x}_j$ to $\mathbf{x}_{j-1}$ under a fixed protocol value $\lambda_{j-1}$. The corresponding forward transition occurs from $\mathbf{x}_{j-1}$ to $\mathbf{x}_{j}$ under the same protocol. Then, the probability for the reverse path can be written by using a different discretization-parameter $\gamma$. One find that  
\begin{eqnarray}
\lefteqn{\Pi[\bar{\mathbf{x}}(t);\bar{\lambda}(t)]}\nonumber\\
&=&\lim_{N\to\infty}\mathcal{N}
e^{-\sum_{j=1}^{N}\Delta t\left[\frac{1}{4}\bar{\mathbf{h}}_{N-j+1}^{(\gamma)}\cdot\mathsf{D}^{-1}\cdot\bar{\mathbf{h}}_{N-j+1}^{(\gamma)}+\gamma (\partial_{\mathbf{x}}\cdot\bar{\mathbf{g}})_{N-j+1}\right]}\nonumber\\
&=&\lim_{N\to\infty}\mathcal{N}
e^{-\sum_{j=1}^{N}\Delta t\left[\frac{1}{4}\widetilde{\mathbf{h}}_{j}^{(1-\gamma)}\cdot\mathsf{D}^{-1}\cdot
\widetilde{\mathbf{h}}_{j}^{(1-\gamma)}+\gamma (\partial_{\mathbf{x}}\cdot\mathbf{g})_{j-1}\right]}~,\label{reverse}
\end{eqnarray}
where $\bar{\mathbf{h}}^{(\gamma)}_j$ is given by $\mathbf{h}_j^{(\gamma)}$ in Eq.~(\ref{forward_segment}) with the replacement of $\mathbf{x}_j$ by $\bar{\mathbf{x}}_j$, and    
$\widetilde{\mathbf{h}}_{j}^{(1-\gamma)}=
-(\mathbf{x}_{j}-\mathbf{x}_{j-1})/\Delta t-\mathbf{g}_{j}^{(1-\gamma)}$. 

Using Eqs.~(\ref{forward}) and (\ref{reverse}), one can find the expression for the EP in Eq.~(\ref{detailed_FR}) as 
\begin{eqnarray}
\lefteqn{k_B^{-1}\Delta S_{\textrm{env}}}\nonumber\\
&=&\sum_{j=1}^{N}\left[\Delta\mathbf{x}_{j}\cdot \mathsf{D}^{-1}\cdot\frac{\mathbf{g}_j^{(\alpha)}
+\mathbf{g}_j^{(1-\gamma)}}{2}-(\alpha-\gamma)\Delta t \partial_x\cdot\mathbf{g}_j\right]\nonumber\\
&=&\sum_{j=1}^{N}\left[\Delta\mathbf{x}_j\cdot \mathsf{D}^{-1}\cdot\mathbf{g}_j^{(1/2)}\right.\nonumber\\
&&\left.-(\alpha-\gamma)\left(\Delta t\partial_{\mathbf{x}}\cdot \mathbf{g}_j-\frac{1}{2}\Delta\mathbf{x}_j\cdot\mathsf{D}^{-1}\cdot(\Delta\mathbf{x}_j
\cdot\partial_{\mathbf{x}})\mathbf{g}_j\right)\right] 
\nonumber\\
&=&\sum_{j=1}^{N}\Delta\mathbf{x}_j\cdot \mathsf{D}^{-1}\cdot\mathbf{g}_j^{(1/2)}~,
\end{eqnarray}
where $\Delta\mathbf{x}_j=\mathbf{x}_j-\mathbf{x}_{j-1}$, and higher-order corrections of order $(\Delta t)^{1/2}$ are neglected. Note that $\mathbf{g}_j^{(\alpha)}\simeq\mathbf{g}^{(1/2)} +(\alpha-1/2)(\Delta\mathbf{x}_j\cdot\partial_{\mathbf{x}})\mathbf{g}_j$. Then, one can show
$\Delta\mathbf{x}_j\cdot\mathsf{D}^{-1}\cdot(\Delta\mathbf{x}_j
\cdot\partial_{\mathbf{x}})\mathbf{g}_j=2\textrm{Tr}[(\partial_{\mathbf{x}}\mathbf{g}_j)\mathsf{D}^{-1}(\Delta\mathbf{x}_j\Delta\mathbf{x}_j)]$, 
which goes to $2\Delta t\partial_{\mathbf{x}}\cdot\mathbf{g}_j$ when
using $\Delta\mathbf{x}_j\Delta\mathbf{x}_j=2\Delta t\mathsf{D}$. Therefore, the term with the factor $(\alpha-\gamma)$ vanishes. Note that the EP is uniquely defined by the midpoint representation, irrespective of the representations used for the forward- and the reverse-path probabilities. The importance of the midpoint representation for the EP lies in the fact that $\mathbf{g}_j^{(1/2)}$ is the Taylor expansion of $\mathbf{g}_j$ up to $\mathcal{O}(\Delta\mathbf{x}_j)$, so $\Delta\mathbf{x}_j\cdot\mathsf{D}^{-1}\cdot\mathbf{g}_j^{(1/2)}$ can be found up to the second order in $\Delta \mathbf{x}_j$, which is the desired order $\mathcal{O}(\Delta t)$. For a simple example in one dimension, consider $\mathbf{g}=-kx\mathbf{i}$. Then, $\Delta S_{\textrm{env}}$ is proportional to $(x_j-x_{j-1})kx_j^{(1/2)}=k(x_j-x_{j-1})(x_j+x_{j-1})/2=k(x_j^2-x_{j-1}^2)/2$, which becomes $\mathcal{O}(\Delta t)$. 

In the continuous-time limit, one can write $dS_{\textrm{env}}/dt=k_B\dot{\mathbf{x}}\cdot\mathsf{D}^{-1}\cdot\mathbf{g}$. The Fokker-Planck equation, Eq.~(\ref{FP_equation}), can be written as
$\partial_t\rho=-\partial_{\mathbf{x}}\cdot\mathbf{j}$,  with the probability current $\mathbf{j}=(\mathbf{g}-\mathsf{D}\cdot\partial_{\mathbf{x}})\rho$. Let the steady-state PDF be $\rho_{\textrm{ss}}\propto e^{-\beta\phi}$ with a potential $\phi(\mathbf{x})$. Then, the steady-state current is given by $\mathbf{j}_{\textrm{ss}}=(\mathbf{g}+\beta\mathsf{D}\cdot\partial_{\mathbf{x}}\phi)\rho_{\textrm{ss}}$. The detailed balance (DB) was found to be characteristic of equilibrium and to be measured by $\mathbf{j}_{\textrm{ss}}$~\cite{kurchan,kwon-ao-thouless,kwon-ao}. The DB holds if $\mathbf{g}=-\beta\mathsf{D}\cdot\partial_{\mathbf{x}}\phi$ while it breaks if $\mathsf{D}^{-1}\cdot\mathbf{g}$ has a non-conservative part  $\mathbf{f}_{\textrm{nc}}$. Generally, in the presence of a time-dependent protocol, the drift term can be written as 
\begin{equation}
\mathbf{g}=\beta\mathsf{D}\cdot\left[-\partial_{\mathbf{x}}\phi(\mathbf{x},\lambda(t))+\mathbf{f}_{\textrm{nc}}(\mathbf{x},t)\right]~,
\end{equation} 
where the time-dependent protocol is supposed to change the potential; otherwise it can be included in $\mathbf{g}_{\textrm{nc}}$. $(\beta\mathsf{D})^{-1}\cdot\mathbf{g}$ can be said to play the role of the force exerted on a particle in motion. From this point of view, one can write the generalized Clausius law as
\begin{equation}
\frac{dS_{\textrm{env}}}{dt}=\frac{1}{T}\left[-\frac{d\phi}{dt}+\dot{W}\right]=\frac{\dot{Q}}{T}~,
\label{Clausius}
\end{equation}
where $d\phi/dt=\partial_{\mathbf{x}}\phi\cdot\dot{\mathbf{x}}+(\partial\phi/\partial\lambda)\dot{\lambda}$ has been used.
$\dot{W}$ is interpreted as the rate of work produced by the nonequilibrium sources, and is given by
\begin{equation}
\dot{W}=\mathbf{f}_{\textrm{nc}}\cdot\dot{\mathbf{x}}+\frac{\partial\phi}{\partial\lambda}\dot{\lambda}~.
\end{equation}
$\dot{Q}$ is interpreted as the rate of heat produced in the heat reservoir, which implies the generalized first law of thermodynamics, 
$d\phi/dt=\dot{W}-\dot{Q}$. One can envisage a stochastic process governed by the Langevin equation as a thermodynamic process, which maybe is out of equilibrium.

In the overdamped limit for the Brownian motion of a particle described in (ii) in the Introduction, one can neglect the inertia part $\dot{\mathbf{p}}$ in Eq.~(\ref{underdamped}) and finds
$\dot{x}=\mathsf{G}^{-1}\cdot\mathbf{f}+\mathsf{G}^{-1}\cdot\boldsymbol{\zeta}$.
Then, one finds $\mathbf{g}=\mathsf{G}^{-1}\cdot\mathbf{f}$ and $\mathsf{D}=\mathsf{G^{-1}S(G^{-1})^t}$. For the heat reservoir in equilibrium at temperature $T$, the Einstein relation $\mathsf{S}=\beta^{-1}\mathsf{G}$ holds so that one can find $\dot{\mathbf{x}}\cdot\mathsf{D}^{-1}\cdot\mathbf{g}=\beta\dot{\mathbf{x}}\cdot\mathbf{f}$. Identifying $\mathbf{f}$ as $-\partial_{\mathbf{x}}\phi(\mathbf{x},\lambda(t))+\mathbf{f}_{\textrm{nc}}$, the above result in Eq.~(\ref{Clausius}) has a physical meaning. 

\section{Brownian motion of a single particle}

The differential equation for the PDF associated with Eq.~(\ref{underdamped}) is given by the Kramers equation~\cite{risken}, which is the Fokker Planck equation for $\rho(\mathbf{x},\mathbf{p})$: 
\begin{equation}
\partial_t \rho(\mathbf{x},\mathbf{p},t)=-H_{\textrm{K}}\rho(\mathbf{x},\mathbf{p},t)
\end{equation}
with
\begin{equation}
H_{\textrm{K}}=\partial_{\mathbf{x}}\cdot\frac{\mathbf{p}}{m}
+\partial_{\mathbf{p}}\cdot\left(-\mathsf{G}\cdot\frac{\mathbf{p}}{m}+\mathbf{f}-\mathsf{S}\cdot\partial_{\mathbf{p}}\right)~,
\end{equation}
where $\partial_{\mathbf{p}}=\boldsymbol{\nabla}_{\mathbf{p}}$.
Using the quantum-mechanical formalism, it can be rewritten as 
\begin{equation}
H_{\textrm{K}}=i\hat{\mathbf{a}}\cdot\frac{\mathbf{p}}{m}+i\hat{\mathbf{b}}\cdot\left(-\mathsf{G}\cdot\frac{\mathbf{p}}{m}+\mathbf{f}\right)+\hat{\mathbf{b}}\cdot\mathsf{S}\cdot\hat{\mathbf{b}}~,
\end{equation}
where $\hat{\mathbf{a}}=i^{-1}\partial_{\mathbf{x}}$ is the same operator as in the last section and $\hat{\mathbf{b}}=i^{-1}\partial_{\mathbf{p}}$ is the conjugate operator for $\mathbf{p}$. 
Using the commutation relation,  $[\hat{\mathbf{b}},\mathbf{f}]=i^{-1}\partial_{\mathbf{p}}\mathbf{f}$, one can write
\begin{equation}
H_{\textrm{K}}=(1-\alpha)H_{\textrm{K}}+\alpha \tilde{H}_{\textrm{K}},
\end{equation}
where 
\begin{eqnarray}
\tilde{H}_{\textrm{K}}&=&i\frac{\mathbf{p}}{m}\cdot\hat{\mathbf{a}}+i\left(-\frac{\mathbf{p}}{m}\cdot\mathsf{G}^{\textrm{t}}+\mathbf{f}\right)\cdot\hat{\mathbf{b}}
+\hat{\mathbf{b}}\cdot\mathsf{S}\cdot\hat{\mathbf{b}}\nonumber\\
&&-\frac{\textrm{Tr}\mathsf{G}}{m}
+\partial_{\mathbf{p}}\cdot\mathbf{f}~.
\end{eqnarray}
Let $|\mathbf{a}\mathbf{b}\rangle$ be a simultaneous eigenstate of $\hat{\mathbf{a}}$ and $\hat{\mathbf{b}}$ that has the eigenfunction given by $\langle\mathbf{x}\mathbf{p}|\mathbf{a}\mathbf{b}\rangle=(2\pi)^{-d}e^{i\mathbf{a}\cdot\mathbf{x}+i\mathbf{b}\cdot\mathbf{p}}$.

In the same time-discretization scheme as in the last section, the path probability can be found for a discrete path $\{ 0 \le j\le N|\mathbf{x}_j,\mathbf{p}_j\}$. Using the $\alpha$-representation, one can find
\begin{eqnarray}
\lefteqn{\Pi[\mathbf{x}(t),\mathbf{p}(t);\lambda(t)]=\lim_{N\to\infty}\prod_{j=1}^{N}\langle\mathbf{x}_j\mathbf{p}_j|e^{-\Delta t H_{\textrm{K}}}|\mathbf{x}_{j-1}\mathbf{p}_{j-1}\rangle}\nonumber\\
&=&\lim_{N\to\infty}\prod_{j=1}^{N}\Big[\langle\mathbf{x}_j\mathbf{p}_j| \mathbf{a}\mathbf{b}\rangle\langle \mathbf{a}\mathbf{b}|\left[1-\Delta t(1-\alpha)H_{\textrm{K}} \right]|\mathbf{x}_{j-1}\mathbf{x}_{j-1}\rangle
\nonumber\\
&&\left.-\langle\mathbf{x}_j\mathbf{p}_j|\alpha\Delta t \tilde{H}_{\textrm{K}}| \mathbf{a}\mathbf{b}\rangle\langle \mathbf{a}\mathbf{b}|\mathbf{x}_{j-1}\mathbf{p}_{j-1}\rangle\right]
\nonumber\\
&=&\lim_{N\to\infty}\prod_{j=1}^{N}\int \frac{d\mathbf{a}}{(2\pi)^d} e^{i\mathbf{a}\cdot[(\mathbf{x}_j-\mathbf{x}_{j-1})-\Delta t\mathbf{p}_j^{(\alpha)}/m]} \nonumber\\
&&\times\int \frac{d\mathbf{b}}{(2\pi)^d}
e^{-\Delta t\mathbf{b}\cdot\mathsf{S}\cdot\mathbf{b}
+i\mathbf{b}\cdot(\mathbf{p}_j-\mathbf{p}_{j-1})-i\Delta t\mathbf{b}\cdot (-\mathbf{G}\cdot\mathbf{p}_j^{(\alpha)}/m+\mathbf{f}_j^{(\alpha)})} \nonumber\\
&&\times e^{\Delta t\alpha[(\textrm{Tr}\mathsf{G})/m-\partial_{\mathbf{p}}\cdot\mathbf{f}_j]} 
  \nonumber\\
&=&\lim_{N\to\infty}\prod_{j=1}^{N}\frac{\delta(\mathbf{x}_j-\mathbf{x}_{j-1}-\Delta t\mathbf{p}_j^{(\alpha)}/m)}{[\textrm{det}(4\pi\Delta t\mathsf{S})]^{1/2}}\nonumber\\
&&\times e^{-\Delta t\sum_{j=1}^{N}[\mathbf{r}_j^{(\alpha)}\cdot\mathsf{S}^{-1}\cdot\mathbf{r}_j^{(\alpha)}/4
-\alpha(\textrm{Tr}\mathsf{G}/m-\partial_{\mathbf{p}}\cdot\mathbf{f}_j)]}~,
\label{path_prob_brownian}
\end{eqnarray}
where $\mathbf{r}_j^{(\alpha)}=(\mathbf{p}_j-\mathbf{p}_{j-1})/\Delta t+\mathsf{G}\cdot\mathbf{p}_j^{(\alpha)}/m-\mathbf{f}_j^{(\alpha)}$. In this expression,  $\mathbf{f}_j^{(\alpha)}=(1-\alpha)\mathbf{f}_{j-1}+\alpha\mathbf{f}_j$ is an intermediate value of force $\mathbf{f}_k=\mathbf{f}(\mathbf{x}_k,\mathbf{p}_k,\lambda_{j-1})$ for $k=j-1$ and $j$, where the protocol has the fixed value at the earlier time. In the continuous-time limit, Eq.~(\ref{path_prob_brownian}) can be written as $\prod_t\delta(\dot{\mathbf{x}}-\mathbf{p}/m) e^{-\int_{0}^{\tau} dt[\mathbf{r}\cdot\mathsf{S}^{-1}\cdot\mathbf{r}/4-\alpha(\textrm{Tr}\mathsf{G}/m
-\partial_{\mathbf{p}}\cdot\mathbf{f})]}$ up to the normalization constant.

The corresponding time-reverse path is given by $\{0\le j\le N|\bar{\mathbf{x}}_j=\mathbf{x}_{N-j},\bar{\mathbf{p}}_j=-\mathbf{p}_{N-j} \}$. For a time interval between $t_{j-1}$ and $t_j$, the force is denoted by $\bar{\mathbf{f}}_k=\mathbf{f}(\bar{\mathbf{x}}_k,\bar{\mathbf{p}}_k,\lambda_{j})$ for $k=j-1$ and $j$, where the protocol has the later time. Replacing $j$ by $N-j+1$, the transition occurs reversely from $(\mathbf{x}_j,-\mathbf{p}_j)$ to $(\mathbf{x}_{j-1},-\mathbf{p}_{j-1})$ under a fixed protocol $\lambda_{j-1}$. The corresponding forward transition occurs from $(\mathbf{x}_{j-1},\mathbf{p}_{j-1})$ to $(\mathbf{x}_j,\mathbf{p}_j)$ under the same protocol. The probability for the time-reverse path can be found in $\gamma$-representation as
\begin{eqnarray}
\lefteqn{\Pi[\bar{\mathbf{x}}(t),\bar{\mathbf{p}}(t);\bar{\lambda}(t)]}
\nonumber\\
&=&\lim_{N\to\infty}\prod_{j=1}^{N}\frac{\delta(\bar{\mathbf{x}}_{N-j+1}-\bar{\mathbf{x}}_{N-j}-\Delta t\bar{\mathbf{p}}_{N-j+1}^{(\gamma)}/m)}{[\textrm{det}(4\pi\Delta t\mathsf{S})]^{1/2}}
\nonumber\\
&&\times e^{-\Delta t\sum_{j=1}^{N}[\bar{\mathbf{r}}_{N-j+1}^{(\gamma)}\cdot\mathsf{S}^{-1}\cdot
\bar{\mathbf{r}}_{N-j+1}^{(\gamma)}/4-\gamma(\textrm{Tr}\mathsf{G}/m
-\partial_{\bar{\mathbf{p}}}\cdot\bar{\mathbf{f}}_{N-j+1})]}
\nonumber\\
&=&\lim_{N\to\infty}\prod_{j=1}^{N}\frac{\delta(\mathbf{x}_j-\mathbf{x}_{j-1}-\Delta t\mathbf{p}_j^{(1-\gamma)}/m)}{[\textrm{det}(4\pi\Delta t\mathsf{S})]^{1/2}}\nonumber\\
&&\times e^{-\Delta t\sum_{j=1}^{N}[\widetilde{\mathbf{r}}_{j}^{(\gamma)}\cdot\mathsf{S}^{-1}\cdot
\widetilde{\mathbf{r}}_{j}^{(\gamma)}/4-\gamma(\textrm{Tr}\mathsf{G}/m
-\partial_{\mathbf{p}}\cdot\widetilde{\mathbf{f}}_{j-1})]}~,
\end{eqnarray}
where $\bar{\mathbf{r}}_j^{(\gamma)}$ is given by $\mathbf{r}_j^{(\gamma)}$ with the replacement: $\mathbf{x}_j, \mathbf{p}_j\to \bar{\mathbf{x}}_j,\bar{\mathbf{p}}_j$. Then one finds $\widetilde{\mathbf{r}}_{j}^{(\gamma)}=(\mathbf{p}_{j}-\mathbf{p}_{j-1})/\Delta t-\mathsf{G}\cdot\mathbf{p}_{j}^{(1-\gamma)}/m-\widetilde{\mathbf{f}}_{j}^{(1-\gamma)}$ for $\widetilde{\mathbf{f}}_j=\mathbf{f}(\mathbf{x}_j,-\mathbf{p}_j,\lambda_j)$. Here, $\mathbf{p}_{j-1}^{(\gamma)}=\mathbf{p}_{j}^{(1-\gamma)}$ has been used.

It is useful to write $\mathbf{f}^{\textrm{rev}}=(\mathbf{f}+\widetilde{\mathbf{f}})/2$ and 
$\mathbf{f}^{\textrm{irr}}=(\mathbf{f}-\widetilde{\mathbf{f}})/2$. Under $\mathbf{p}\to-\mathbf{p}$, the former is symmetric while the latter antisymmetric. 
Write $\Delta\mathbf{p}_j=\mathbf{p}_j-\mathbf{p}_{j-1}$. Then, the EP in Eq.~(\ref{detailed_FR}) can be written as
\begin{eqnarray}
\lefteqn{k_B^{-1}\Delta S_{\textrm{env}}}\nonumber\\
&=&-\frac{1}{2}\sum_{j=1}^{N\to\infty}
\left[\left(\Delta\mathbf{p}_j-\Delta t\mathbf{f}_j^{\textrm{rev}(\alpha)}\right)\cdot \mathsf{S}^{-1}\cdot\left(\mathsf{G}\cdot\frac{\mathbf{p}_j^{(\alpha)}}{m}-\mathbf{f}_j^{\textrm{irr}(\alpha)}\right)\right.
\nonumber\\
&&+\left(\Delta\mathbf{p}_j-\Delta t\mathbf{f}_j^{\textrm{rev}(1-\gamma)}\right)\cdot \mathsf{S}^{-1}\cdot\left(\mathsf{G}\cdot\frac{\mathbf{p}_j^{(1-\gamma)}}{m}-\mathbf{f}_j^{\textrm{irr}(1-\gamma)}\right)
\nonumber\\
&&-\Delta\mathbf{p}_j\cdot\mathsf{S}^{-1}\cdot\left(\mathbf{f}_j^{\textrm{rev}(\alpha)}-\mathbf{f}_j^{\textrm{rev}(1-\gamma)}\right)
\nonumber\\
&&-2\Delta t\left( \frac{\alpha-\gamma}{m}\textrm{Tr}\mathsf{G}-\partial_{\mathbf{p}}\cdot(\alpha\mathbf{f}_j
+\gamma\widetilde{\mathbf{f}}_j)\right)\Bigg]~.
\end{eqnarray}
One should keep the terms in the bracket up to $\mathcal{O}(\Delta t)$ and note that the outer product $\Delta\mathbf{p}_j\Delta\mathbf{p}_j$ is equivalent to  $2\Delta t\mathsf{S}$. One can expand $\mathbf{f}_j^{(\alpha)}=\mathbf{f}_j^{(1/2)}+(\alpha-1/2)
(\Delta \mathbf{p}_j\cdot\partial_{\mathbf{p}})\mathbf{f}_j$ and do the same for the other vectors. After rearranging the terms, one can simplify the EP in the time interval between $t_{j-1}$ and $t_{j}$ as
\begin{eqnarray}
k_B^{-1}\Delta S_{\textrm{env,j}}&=&
-\Delta \mathbf{p}_j\cdot\mathsf{S}^{-1}\mathsf{G}\cdot\frac{\mathbf{p}_j^{(1/2)}}{m}+
\Delta t \mathbf{f}_j\cdot\mathsf{S}^{-1}\mathsf{G}\cdot\frac{\mathbf{p}_j}{m}
\nonumber\\
&&+
\Delta\mathbf{p}_j\cdot\mathsf{S}^{-1}\cdot\mathbf{f}_j^{\textrm{irr}(1/2)} 
-\Delta t\mathbf{f}_j^{\textrm{irr}}\cdot\mathsf{S}^{-1}\mathsf{G}
\cdot\frac{\mathbf{p}_j}{m}
\nonumber\\
&&-\Delta t \mathbf{f}_j^{\textrm{rev}}\cdot\mathsf{S}^{-1}
\cdot\mathbf{f}_j^{\textrm{irr}}
-\Delta t\partial_{\mathbf{p}}\cdot\mathbf{f}_j^{\textrm{rev}}
\nonumber\\
&&+\textrm{remainder}~,
\end{eqnarray}
where the term `remainder' is given by
\begin{eqnarray*}
&&(\alpha-\gamma)\left( \Delta t\frac{\textrm{Tr}\mathsf{G}}{m}
-\frac{1}{2}\Delta\mathbf{p}_j\cdot\mathsf{S^{-1}G}\cdot(\Delta\mathbf{p}_j
\cdot\partial_{\mathbf{p}})\frac{\mathbf{p}_j}{m}\right)
\nonumber\\
&&+\frac{1}{2}\Delta\mathbf{p}_j\cdot\mathsf{S}^{-1}\cdot(\Delta\mathbf{p}_j\cdot
\partial_{\mathbf{p}})
\left((\alpha-\gamma)\mathbf{f}_j^{\textrm{irr}}
+(\alpha+\gamma)\mathbf{f}_j^{\textrm{rev}}\right)
\nonumber\\
&&-\Delta t\partial_{\mathbf{p}}\cdot \left((\alpha+\gamma)\mathbf{f}_j^{\textrm{rev}}
+(\alpha-\gamma)\mathbf{f}_j^{\textrm{irr}}\right)~,
\end{eqnarray*}
which can be shown to vanish by using $\Delta\mathbf{p}_j\cdot\mathsf{S^{-1}G}\cdot(\Delta\mathbf{p}_j\cdot
\partial_{\mathbf{p}})\mathbf{p}_j=2\textrm{Tr}\mathsf{G}/m$ and $\Delta\mathbf{p}_j\cdot\mathsf{S}^{-1}\cdot(\Delta\mathbf{p}_j\cdot
\partial_{\mathbf{p}})\mathbf{f}_j=2\partial_{\mathbf{p}}\cdot\mathbf{f}_j$. Therefore, the EP is uniquely defined
from the midpoint representation, independent of the specific representations used for the path probabilities. 
Note that only the terms coupled with $\Delta\mathbf{p}_j$ should be written in the midpoint representation.  

In the continuous-time limit, the EP can be written as
\begin{equation}
\frac{dS_{\textrm{env}}}{dt}=\textrm{Tr}\dot{\mathsf{Q}}\mathsf{T}^{-1}+\frac{dS_{\textrm{an}}}{dt}~.
\end{equation}
The matrix $\mathsf{T}^{-1}$ of the reciprocal temperature is defined as $k_B\mathsf{S^{-1}G}$. The matrix $\dot{\mathsf{Q}}$ of the heat production rate is defined as $-(\dot{\mathbf{p}}-\mathbf{f})\mathbf{p}/m$, the trace of which is the real rate of heat produced in the heat reservoir. 
The second term is the rate of the anomalous EP, which was found in a recent work~\cite{kwon-yeo-lee-park}. One can get
\begin{eqnarray}
\textrm{Tr}\dot{\mathsf{Q}}\mathsf{T}^{-1}&=&-(\dot{\mathbf{p}}-\mathbf{f})\cdot\mathsf{S^{-1}G}\cdot\frac{\mathbf{p}}{m}~,
\label{EP_Brown}\\
\frac{dS_{\textrm{an}}}{dt}&=&
\mathbf{f}^{\textrm{irr}}\cdot\mathsf{S}^{-1}\cdot\left(\dot{\mathbf{p}}-
\mathsf{G}\cdot\frac{\mathbf{p}}{m}-\mathbf{f}^{\textrm{rev}}\right)
-\partial_{\mathbf{p}}\cdot\mathbf{f}^{\textrm{rev}} ~.\label{anomalous_EP}
\end{eqnarray}
Here, the midpoint representation is implicit while it should be explicit for a rigorous calculation of the EP. 

Equation~(\ref{EP_Brown}) gives the expression of the EP for a general heat reservoir that might not be in equilibrium for asymmetric $\mathsf{T}^{-1}=k_B\mathsf{S^{-1}G}$.  The Brownian motion of a colloidal particle in a liquid, where the liquid plays the role of an equilibrium heat reservoir at a fixed temperature $T$, has been of much interest~\cite{wang,trepagnier,hummer,pak}. In this case, one has $\mathsf{G}=\gamma\mathsf{I}$ and $\mathsf{T}=T\mathsf{I}$, and finds $\mathsf{S}=\gamma k_BT\mathsf{I}$, which is known as the Einstein relation or the fluctuation-dissipation relation. Therefore,  one can find  $\textrm{Tr}\dot{\mathsf{Q}}\mathsf{T}^{-1}=-(\dot{\mathbf{p}}-\mathbf{f})\cdot\mathbf{p}/{mT}
=\dot{Q}/T$, where $\dot{Q}$ is the rate of work extracted by the force, $-\gamma\mathbf{p}/m+\boldsymbol{\xi}$, exerted by the reservoir, which is, by definition, the heat production rate. One can write $\mathbf{f}=-\partial_{\mathbf{x}}V(\mathbf{x},\lambda(t))+\mathbf{f}_{\textrm{nc}}(\mathbf{x},\mathbf{p})$. In the discrete-time representation, one can find
\begin{eqnarray}
\lefteqn{-(\dot{\mathbf{p}}-\mathbf{f})\cdot\frac{\mathbf{p}}{m}}\nonumber\\
&=& -\frac{1}{m}\frac{\Delta\mathbf{p}_j}{\Delta t}\cdot\mathbf{p}_j^{(1/2)}
+\left(-\partial_{\mathbf{x}}V_j+\mathbf{f}_{nc,j}\right)\cdot\frac{\mathbf{p}_j}{m}
\nonumber\\
&=&-\frac{\mathbf{p}^2_j-\mathbf{p}^2_{j-1}}{2m\Delta t}
-\frac{dV}{dt}+\frac{\partial V}{\partial\lambda}\dot{\lambda}+\mathbf{f}_{nc,j}\cdot\frac{\mathbf{p}_j}{m}
\nonumber\\
&=&-\frac{d E}{dt}+\dot{W}=\dot{Q}~,
\label{1st_law}
\end{eqnarray}  
where $E=\mathbf{p}^2/(2m)+V(\mathbf{x},\lambda)$ is the total energy and $\dot{W}$ is the work production rate, which is composed of two contributions:
\begin{equation}
\dot{W}=\frac{\partial V}{\partial\lambda}\dot{\lambda}+\mathbf{f}_{nc}\cdot\dot{\mathbf{x}}~.
\end{equation}
The first term was recognized by Jarzynski~\cite{jarzynski}, and the second term is the rate of work done by the nonconservative force. The final line in Eq.~(\ref{1st_law}) is consistent with the first law of thermodynamics. If there is no time-dependent protocol and no nonconservative force, only heat is dissipated by the amount of energy change. In the steady state, therefore, no heat is produced, which is characteristics of equilibrium. On the other hand, if work production occurs from a time-dependent protocol or a nonconservative force, persistent productions of work and heat can occur even in the steady state with $\langle dE/dt\rangle=0$, which is a fundamental difference between an equilibrium steady state and a nonequilibrium steady state.  In the absence of a momentum-dependent force, the Clausius law is, indeed, satisfied, $d S_{\textrm{env}}/dt=Q/T$. However, in the presence of a momentum-dependent force, an anomalous production $\Delta S_{\textrm{an}}$ added to $Q/T$ appears. An anomalous feature is even more serious if an irreversible part $\mathbf{f}^{\textrm{irr}}$ exists, which can be seen in Eq.~(\ref{anomalous_EP}). There is yet no clear understanding of $\Delta S_{\textrm{an}}$ beyond its existence.

\section{Conclusions}

A rigorous expression for the EP, the detailed fluctuation relation, that is composed of important thermodynamic quantities such as energy, work, and heat, is derived. The result is shown to be uniquely given by the midpoint representation independent of any representations used for the forward- and the reverse-path probabilities.

The exact rule for the discrete-time representation is very important in the calculation via the path-integral. In order to find the average value of a thermodynamic quantity, one needs to carry out the path integral with the path probability. Then, one can use any representation for the path probability provided in Eq.~(\ref{forward}) or (\ref{path_prob_brownian}). If the quantity to be averaged is path-dependent, such as EP, work, and heat, it should be expressed in the midpoint representation. For example, $\Delta\mathbf{p}_j\cdot\mathbf{p}_j^{(\alpha)}/m$ goes to the correct kinetic energy change $(\mathbf{p}^2_j-\mathbf{p}^2_{j-1})/(2m)$ only if $\alpha=1/2$ is used. If a different value of $\alpha$ is used, the correction is $(\alpha-1/2)\Delta\mathbf{p}_j\cdot(\Delta\mathbf{p}_j\cdot\partial_{\mathbf{p}})\mathbf{p}_j/m
=2(\alpha-1/2)\Delta t (\textrm{Tr}\mathsf{S})/m$, which is not negligible. The second law of thermodynamics can be expressed by using the positivity of $\langle \dot{S}_{\textrm{tot}}
\rangle=\int d\mathbf{x}d\mathbf{p} \rho(\mathbf{x},\mathbf{p})^{-1} \mathbf{j}^{\textrm{irr}}\cdot\mathsf{S}^{-1}\cdot\mathbf{j}^{\textrm{irr}}$, where $\mathbf{j}^{\textrm{irr}}=(-\mathsf{G}\cdot\mathbf{p}/m+\mathbf{f}^{\textrm{irr}}-
\mathbf{S}\cdot\partial_{\mathbf{p}})\rho$. The midpoint representation of the EP is necessary for the derivation~\cite{kwon-yeo-lee-park}.

The detailed fluctuation relation in Eq.~(\ref{detailed_FR}) is an ingredient to the FT for the total entropy production shown in Eq.~(\ref{FT_EP}). One can also use it to prove other FT's, for example, for work production and heat production. There are two types of the FT. One is called the integral FT (IFT) of type $\langle e^{-R}\rangle=1$ for a time-accumulated quantity $R$. The other is called the detailed FT (DFT) of type $P_{\textrm{F}}(R)/P_{\textrm{R}}(-R)=e^{\beta R}$. The former is a necessary, but not a sufficient, condition for the latter. The IFT can be proven in a way similar to that used for Eq.~(\ref{FT_EP}). The DFT can also be proven by using 
$P_{\textrm{F}}(R)=\langle\delta(R-R[\mathbf{q}(t),\lambda(t)])\rangle$, where the bracket denotes the integral over all paths and the initial state. $P_{\textrm{R}}(-R)$ is given similarly by $\langle\delta(R+R[\mathbf{q}(t),\lambda(t)])\rangle$, where the subscript $\textrm{R}$ stands for the process under time-reverse protocol. See Ref.~\cite{esposito} for the detail.

The investigation in this study is restricted to stochastic motion due to additive noise.  A derivation of the detailed fluctuation relation for multiplicative noise will be presented in the near future. 

\begin{acknowledgements}
This work was supported by the research fund of Myongji University, 2013.
\end{acknowledgements}

\end{document}